# Study of Silicon+$^6$LiF thermal neutron detectors: GEANT4 simulations versus real data

S. Lo Meo[1,2], L. Cosentino[3], P. Bartolomei[1] and P. Finocchiaro[3,*]

[1)] ENEA Bologna, Italy
[2)] INFN Sezione di Bologna, Italy
[3)] INFN Laboratori Nazionali del Sud, Catania, Italy

*) corresponding author

**Abstract**

Research and development on alternative thermal neutron detection technologies and methods are nowadays needed as a possible replacement of $^3$He-based ones. Commercial solid state silicon detectors, coupled with neutron converter layers containing $^6$Li, have been proved to represent a viable solution for several applications as present in literature. In order to better understand the detailed operation and the response and efficiency of such detectors, a series of dedicated GEANT4 simulations were performed and compared with real data collected in a few different configurations. The results show a remarkable agreement between data and simulations, indicating that the behavior of the detector is fully understood.

## 1  Introduction

The lack and the increasing cost of $^3$He have triggered in the last years a worldwide R&D program seeking new techniques for neutron detection. For many applications a realistic alternative is needed to $^3$He-based neutron detectors which so far have been the most widely used systems, as they are almost insensitive to radiation other than thermal neutrons [1],[2],[3].

Several developments involving neutron detection are currently being pursued in the fields of homeland security, nuclear safeguards, nuclear decommissioning and radwaste management. Two possible applications are worth to be mentioned, namely the development of neutron sensitive panels to be placed around nuclear material in a $\approx 4\pi$ solid angle coverage for coincidence neutron counting applications [4], and the deployment of arrays of small neutron detectors for the online monitoring of spent nuclear fuel repositories [5],[6].

In a previous paper [7] it was shown that the use of a fully depleted silicon charged particle detector, in combination with a $^6$LiF neutron converter film, can be successfully exploited to detect thermal neutrons with a reasonable efficiency, as also suggested by other authors [8],[9]. The neutron conversion mechanism is based on the well known reaction

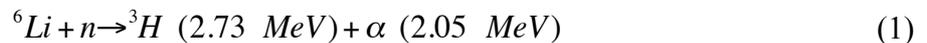

$$^6Li + n \rightarrow {^3H}\ (2.73\ MeV) + \alpha\ (2.05\ MeV) \qquad (1)$$

which is the only possible decay channel following the neutron capture in $^6$Li, and is free of gamma rays. The energy spectrum measured by the silicon detector in such a configuration has a characteristic shape, and allows to discriminate the capture reaction products from the low-energy background basically due to gamma rays.

The feasibility of this technique is indeed well established [10],[11], several applications of the proposed detection technique are already in use, like for instance at the n-TOF spallation neutron beam facility [12],[13], even though a full characterization in terms of response, efficiency and gamma (in)sensitivity has not been performed yet. In this paper such a solid state neutron detector was thoroughly studied by means of simulations, and its response was compared to experimental data taken with a thermalized AmBe neutron source and with a (mostly) thermal neutron beam at the INES facility [14].

## 2   The simulation environment

The simulation code employed for this work is GEANT4 v.10.3 [15]. Even if GEANT4 was originally developed for the high energy physics community, its physics models have been constantly expanding to cover applications at lower energy. In recent years it has been successfully used also to describe the transport of neutrons from thermal energy to GeV energies [16],[17]. The considered solid state detectors were 3 cm x 3 cm double side silicon pads, assembled in four different configurations sketched in Figure 1: a) coupled with a "thin" single layer of $^6$LiF converter (1.6μm thickness); b) coupled with a "thick" single layer of $^6$LiF converter (16 μm thickness); c) coupled with two "thick" layers of $^6$LiF converter (2 x 16 μm thickness) one on each face of the detector; d) stack of two identical samples of the latter (2 silicon detectors and 4 x 16μm $^6$LiF). The $^6$LiF, enriched in $^6$Li at 95%, was deposited onto a 0.6 mm carbon fiber substrate and placed at 1 mm distance from the silicon surface. We remark that the thermal neutron inelastic and capture cross sections on carbon and fluorine are five orders of magnitude lower than on $^6$Li.

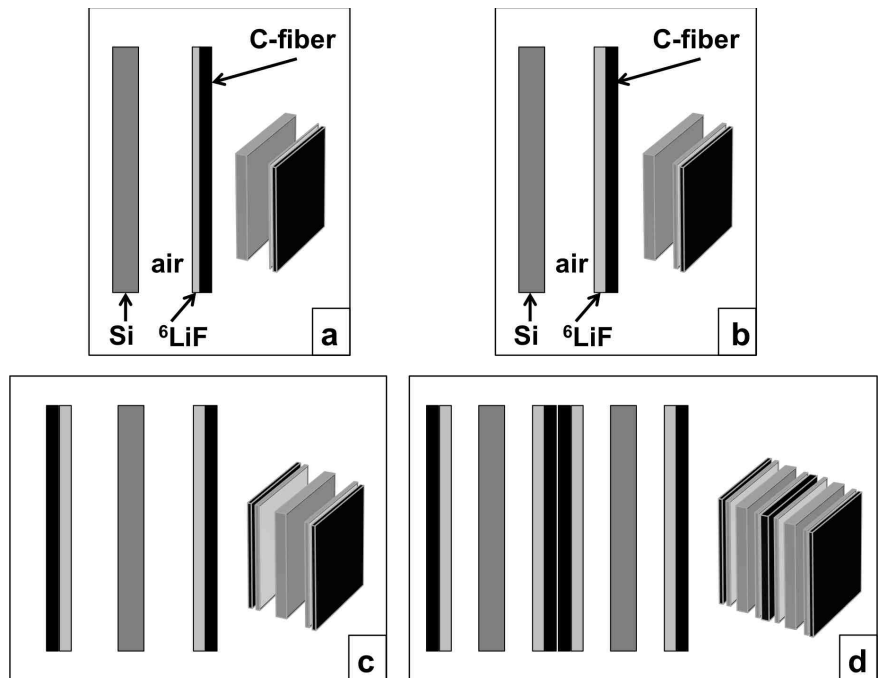

Figure 1. The simulated neutron detector configurations (not to scale), the converter is always deposited on 0.6mm carbon fiber substrate. One silicon detector with a single 1.6μm $^6$LiF converter (a); one silicon detector with a single thicker 16μm $^6$LiF converter (b); one silicon detector with a double 16μm $^6$LiF converter (c); two silicon detectors with four 16μm $^6$LiF converters (d).

The simulated irradiation schemes were basically two: flood, i.e. with a uniform thermal neutron beam perpendicular to the detector face (that is, parallel to the z-axis), and isotropic, i.e. with a uniform thermal neutron flux emitted from a spherical shell surrounding the detector. All of the four detector configurations were simulated with the two irradiation schemes.

As for the experimental data, spectra were available as measured with the four detector configurations. In particular we had data taken with an AmBe neutron source inside a big moderator ($\approx$ 1 m$^3$ volume) thus mimicking the isotropic irradiation, and data taken with a neutron beam at the INES facility mimicking the flood irradiation.

The preliminary check to be done concerned the correctness of the angular distribution of the alphas and tritons produced by the simulation package. Indeed thermal neutron capture in $^6$Li produces alphas and tritons isotropically distributed. Should this not be the case in the simulation, the results could be strongly misleading. Therefore we simulated the interaction of thermal neutrons with the $^6$LiF converter, reporting the direction cosines of each alpha and triton produced. Figure 2 shows the distribution of the three direction cosines for the alphas, that is flat as expected. The same plot for tritons, not shown, is an exact mirror copy of Figure 2 because of the back-to-back emission of alpha and triton. The single cosine distributions are a good indication of the isotropy, but one has to check the combined distributions to make sure that there are no implicit correlations due to the pseudo-random number generator in the simulation. This was done and the result, shown in Figure 3, indicates a satisfactory isotropy as the 3D distribution of the direction cosines fills the unit sphere surface uniformly.

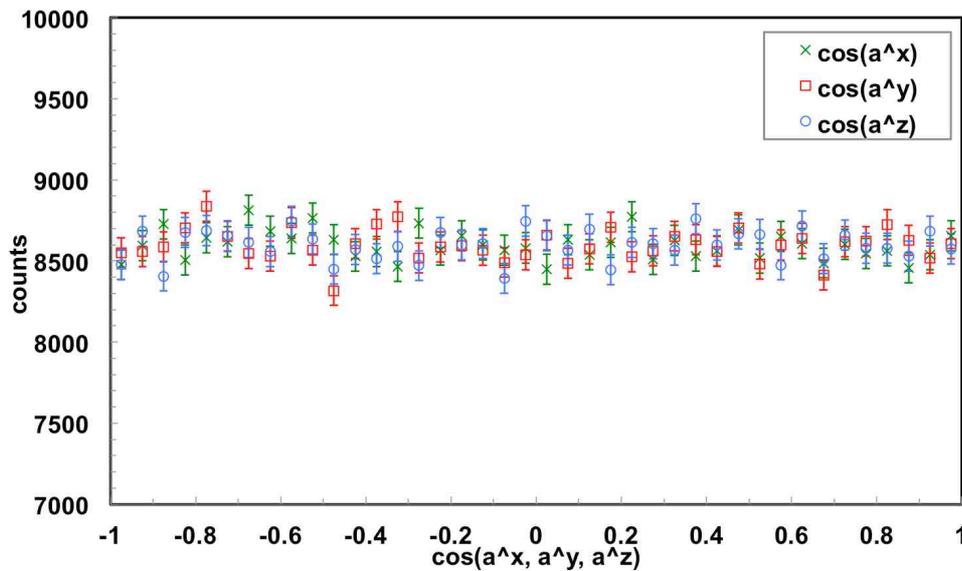

Figure 2. Simulated distribution of the direction cosines of alpha particles produced by thermal neutron capture in $^6$Li. The flat behavior is expected because of the isotropic emission.

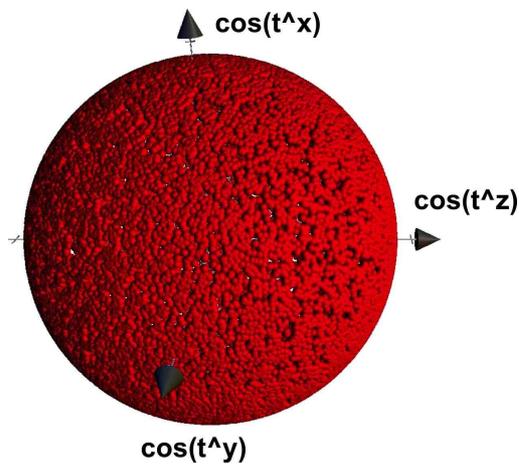

Figure 3. Simulated 3D distribution of the direction cosines of tritons (and alphas) produced by the neutron capture in $^6$Li. The uniform filling of the unit sphere surface indicates that there is no hidden correlation between directions, and that the overall distribution is actually isotropic.

## 3   The reference thin neutron converter

The first detector configuration we simulated is the one with a "thin" 1.6 μm $^6$LiF converter (Figure 1a). In this case the expected spectrum shape is very characteristic, thus allowing to easily disentangle the triton and alpha behavior and to make an immediate comparison with the experimental data. After normalizing for the different number of impinging neutrons due to solid angle, we did not observe a statistically significant difference between the flood and the isotropic results. The distributions of the direction cosines for the alphas and tritons as they leave the converter are shown respectively in Figure 4 and Figure 5. The alpha particles, heavier and with lower energy, are more easily stopped in matter. This is why we observe a non-uniform distribution. In particular, by comparing the z-cosine distribution of alphas and tritons (Figure 6), we see that the alpha particles are constrained in a smaller forward angular region than tritons. Alphas start to be suppressed beyond ≈ 70° [cos(a^z) ≈ 0.35], tritons beyond ≈ 86° [cos(t^z) ≈ 0.075]. Obviously, the detector was placed in the forward direction, therefore the useful part of the plots in Figure 6 for the detection is the positive one.

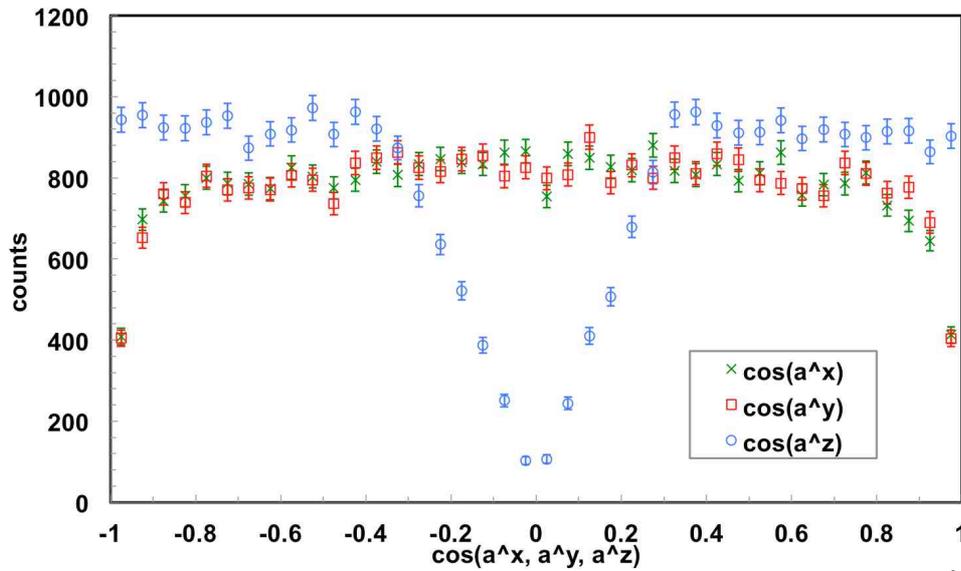

Figure 4. Simulated distribution of the direction cosines of alpha particles leaving the "thin" 1.6μm $^6$LiF converter, indicating a prevalent emission in the forward/backward direction.

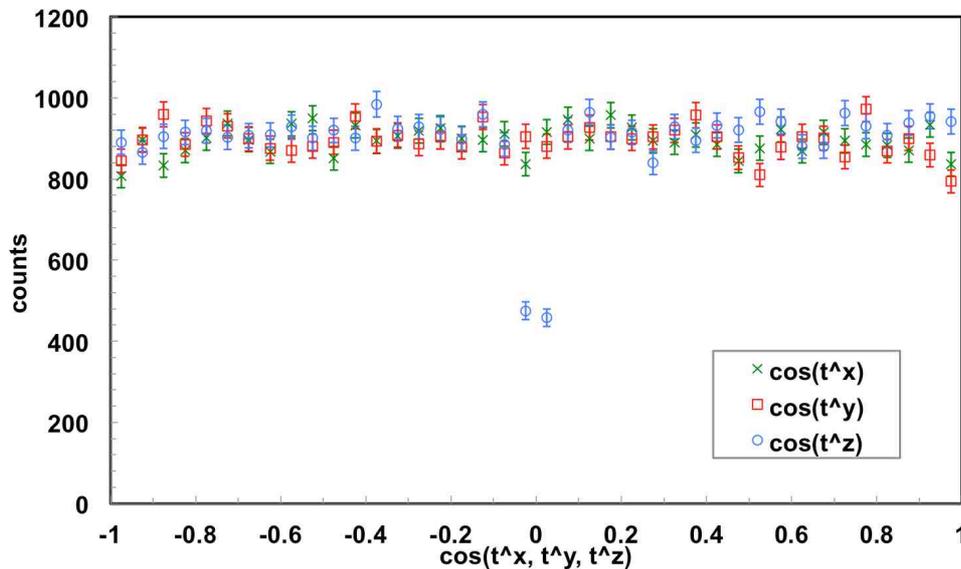

Figure 5. Simulated distribution of the direction cosines of tritons leaving the "thin" 1.6μm $^6$LiF converter, indicating a wide emission with just a slight suppression at large angles.

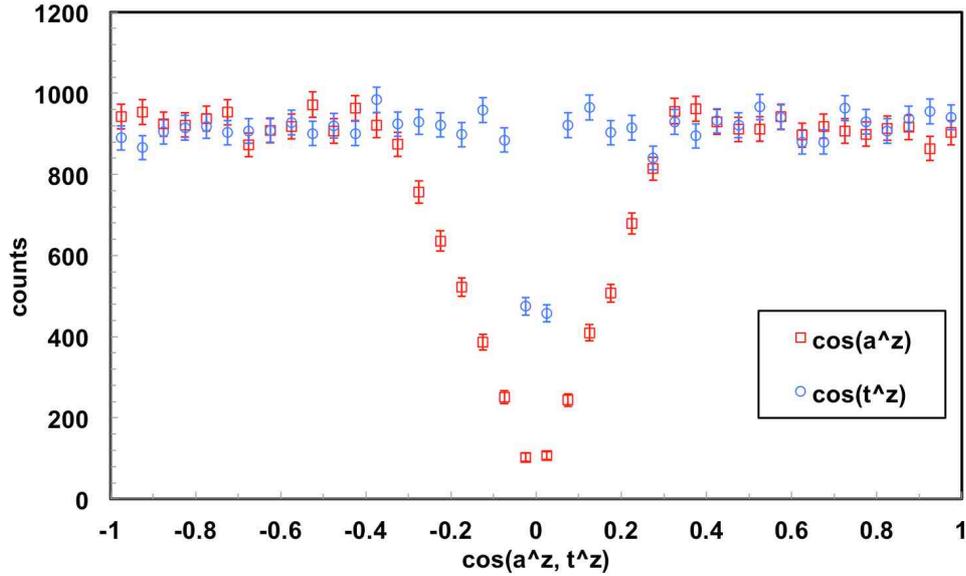

Figure 6. Comparison of the z-direction cosines for alphas and tritons as taken from Figure 4 and Figure 5. As expected, the heavier (and less energetic) alpha particles are constrained in a smaller forward angular region than tritons. Alphas start to be suppressed beyond ≈ 70° [cos(a^z) ≈ 0.35], tritons beyond ≈ 86° [cos(t^z) ≈ 0.075].

In Figure 7 we show the deposited energy spectrum as produced in the flood simulation, along with the separate contributions due to alphas and tritons. The upper endpoint of alphas is clearly seen at 1.9 MeV (we remind that in the reaction (1) alphas are emitted with 2.05 MeV, but there is a 1mm air gap between the converter and the silicon detector). The fraction of tritons below 1.9 MeV is only 3.1%, therefore a detector in this configuration could be reasonably used as a reference for the efficiency calibration of other detectors, using the triton peak area as reference figure. Figure 8 shows again the deposited energy in the flood simulation, as compared with the isotropic simulation and with experimental data collected with the moderated AmBe neutron source. The energy calibration of the silicon detector was done by means of the known alpha and triton upper endpoint energy values. As the neutron flux value inside the moderator was only roughly known, the experimental spectrum had to be rescaled in order to compare it to the simulation results. The scaling factor was chosen in such a way to have the same area of the triton peak. The statistical error bars in the flood spectrum were tiny and were omitted for clarity, and the corresponding spectrum was reported as a continuous line. The isotropic spectrum has a lower statistics and therefore larger error bars, due to the smaller number of neutrons hitting the converter. It was simply normalized to the number of impinging neutrons without further rescaling, and as expected it is in perfect agreement with the flood spectrum. The horizontal error bars shown on the data points actually represent the bin width. They were purposefully chosen larger (in principle one should use the bin width divided by the square root of 12), in order to account for possible systematic uncertainties coming from the energy calibration for the experimental data, and from the pseudo-statistical behavior for the simulated data.

The agreement between data and simulation is quite good down to ≈ 1.3 MeV, where the influence of the huge amount of high energy gamma rays (4.4 MeV) from the AmBe source takes over, as will be discussed in section 7.

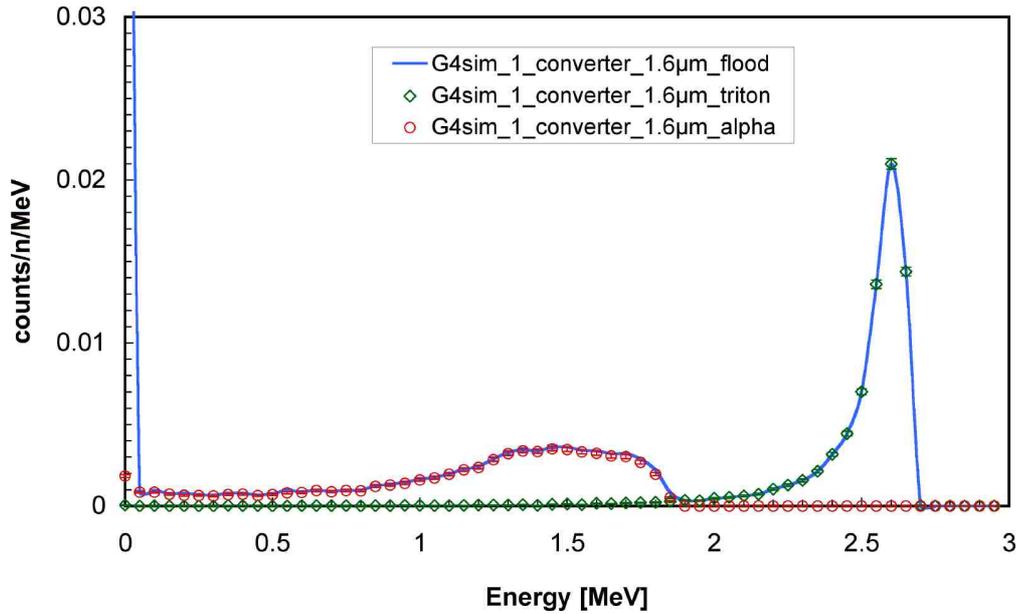

Figure 7. The simulated energy spectrum on the silicon detector for the flood irradiation in the "thin" 1.6 µm $^6$LiF converter configuration (Figure 1a), showing the separate contributions due to alphas and tritons.

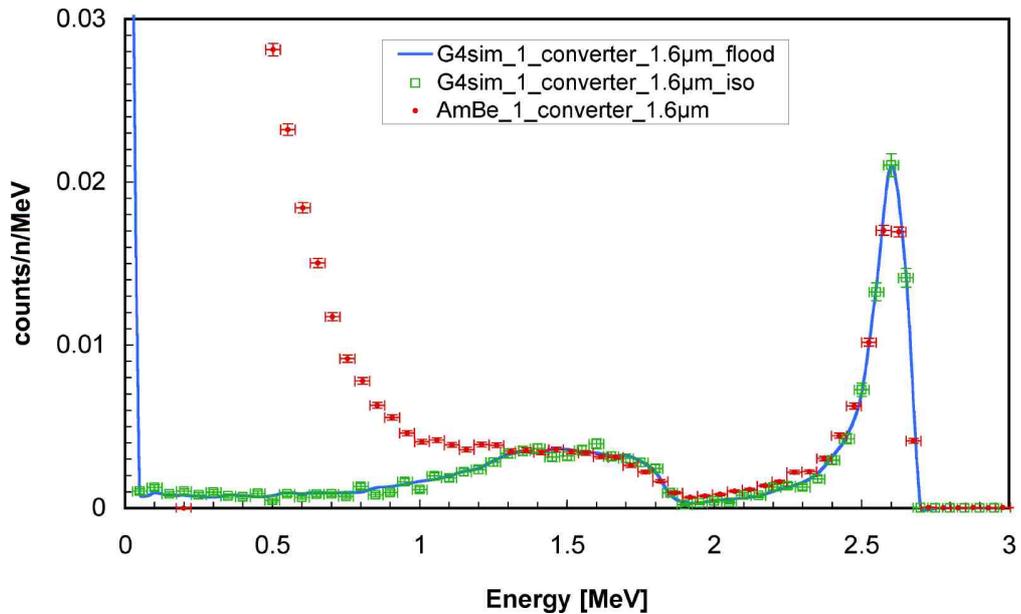

Figure 8. The energy spectrum on the silicon detector for the flood and isotropic simulations in the "thin" 1.6 µm $^6$LiF converter configuration (Figure 1a), compared with an experimental result obtained with a thermalized AmBe neutron source.

## 4  The thick neutron converter

From previous measurements and estimates we were aware that a reasonable energy threshold level to operate the neutron detector should be around 1.5 MeV. This is why the "thick" converter thickness was chosen as 16 µm (Figure 1b): indeed a triton emitted parallel to the z-axis from the deepest part of the converter comes out with 1.63 MeV, therefore still enough to be detected above such a threshold. A thicker converter would make little sense, as it would not be exploited through its full thickness and on the contrary would pose additional issues in terms of its production and mechanical stability.

With such a "thick" converter layer one can expect a relevant reduction of the angular range of emission, which is indeed shown in Figure 9 and Figure 10 where we reported the distribution of the direction cosines respectively for alphas and tritons exiting the converter, in the flood irradiation geometry. Alpha particles are strongly suppressed and show no hint of isotropy, whereas tritons are still isotropically emitted up to 60° [$\cos(t\hat{}z) \approx \pm 0.5$]. This is highlighted in Figure 11, where we compared the z-direction cosines of alphas and tritons. Figure 12 shows the simulated 3D distribution of the direction cosines of alphas (left) and tritons (right) exiting the converter. The strong and slight suppression at large angles, respectively for alphas and tritons, is indicated by the denser distributions around the z-direction. Moreover, in Figure 11 we drew the gridlines to help appreciate a slight systematic tendency of the forward emission to be lower than the backward one, likely due to the ≈ 8% overall absorption of neutrons in the 16μm "thick" converter: more neutron captures occur in the first converter layers than in the last ones. This effect, more pronounced for alphas due to their shorter range in matter, is supported by the corresponding data in the isotropic irradiation geometry which do not show such an asimmetry, as shown in Table 1.

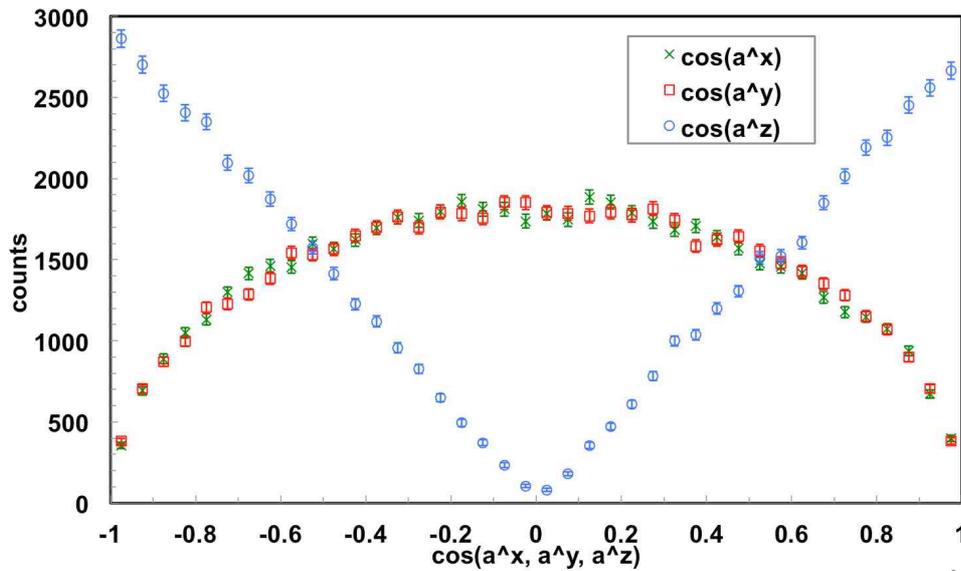

Figure 9. Simulated distribution of the direction cosines of alpha particles leaving the "thick" 16 μm $^6$LiF converter, indicating a relevant suppression with predominant emission in the forward/backward direction.

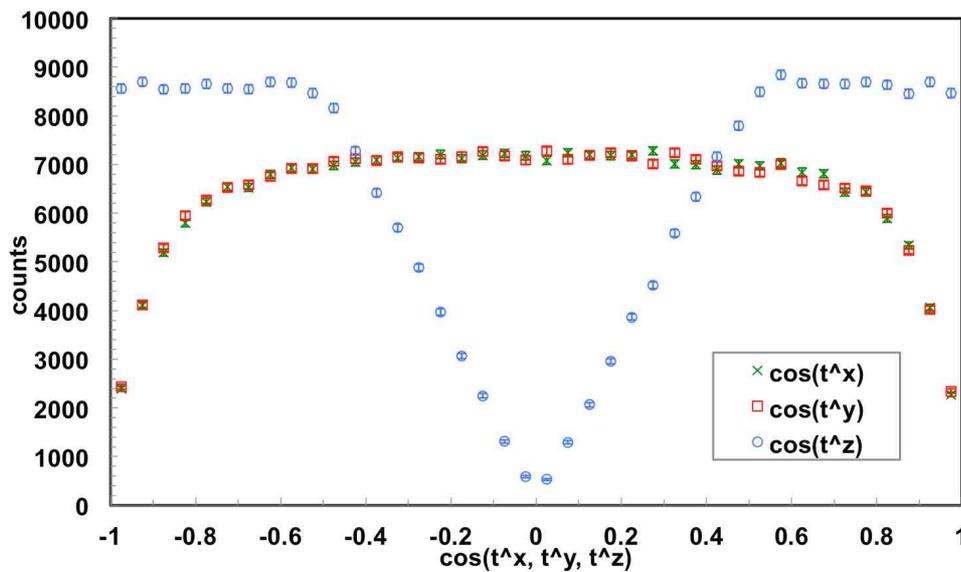

Figure 10. Simulated distribution of the direction cosines of tritons leaving the "thick" 16 μm $^6$LiF converter, indicating a prevalent forward/backward emission with a considerable suppression at large angles.

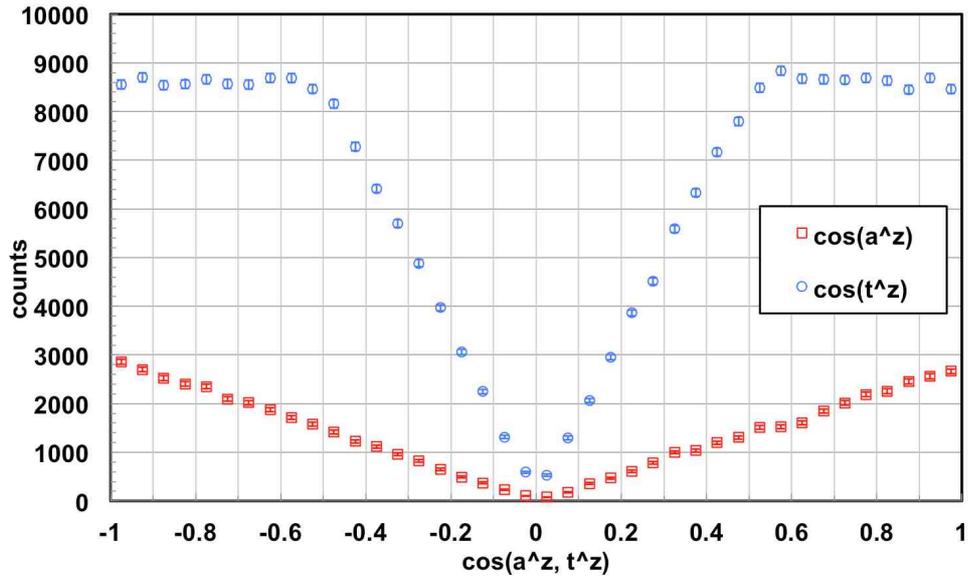

Figure 11. Comparison of the z-direction cosines for alphas and tritons as taken from Figure 9 and Figure 10. As expected, the heavier (and less energetic) alpha particles are strongly suppressed. Tritons start to be suppressed beyond ≈ 60° [cos(t^z) ≈ ±0.5)].

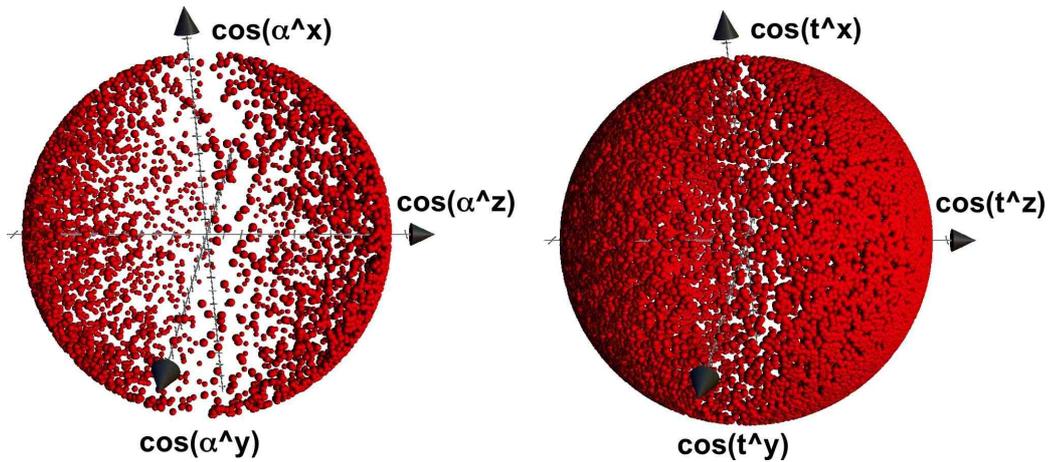

Figure 12. Simulated 3D distribution of the direction cosines of alphas (left) and tritons (right) exiting the "thick" 16µm $^6$LiF converter. The strong and slight suppression at large angles, respectively for alphas and tritons, is indicated by the denser distributions around the z-direction.

|  | forward/backward emission ratio | | | |
| --- | --- | --- | --- | --- |
|  | emission from 16µm $^6$LiF (flood) | deviation from 1 | emission from 16µm $^6$LiF (isotropic) | deviation from 1 |
| alpha | 0.94 ± 0.01 | 6σ | 0.98 ± 0.02 | < 1σ |
| triton | 0.990 ± 0.004 | 2.5σ | 1.008 ± 0.009 | < 1σ |

Table 1. The forward-to-backward ratio for alphas and tritons emitted from the converter in case of flood and isotropic irradiation. The flood data show an asimmetry likely due to neutron absorpion in the "thick" converter.

By simulating an irradiation with a pencil beam in the center of the converter, and reporting the XY coordinates of the alphas and tritons hit points on the silicon detector, one can visually realize the effect of the different angular emission for the two particles (Figure 13). What one expects on this basis is that the geometrical acceptance of the silicon detector should be slightly better for alphas than for tritons, as the tritons produced by the interaction of neutrons close to the converter border are more easily lost to the detection because of their larger angular spread. Indeed, this is what we found, as shown in Table 1. The geometrical efficiency loss due to border effects is rather limited, because of the close proximity of the converter to the silicon detector (1 mm) as compared to the detector size (3 cm x 3 cm).

|  | geometrical acceptance [%] | |
| --- | --- | --- |
|  | total | with E ≥ 1.5 MeV |
| alpha | 96.9 ± 0.5 | 98.5 ± 1.2 |
| triton | 96.3 ± 0.2 | 96.8 ± 0.2 |

Table 2. Geometrical acceptance for alphas and tritons, in the full energy range and in a more realistic E ≥ 1.5 MeV range.

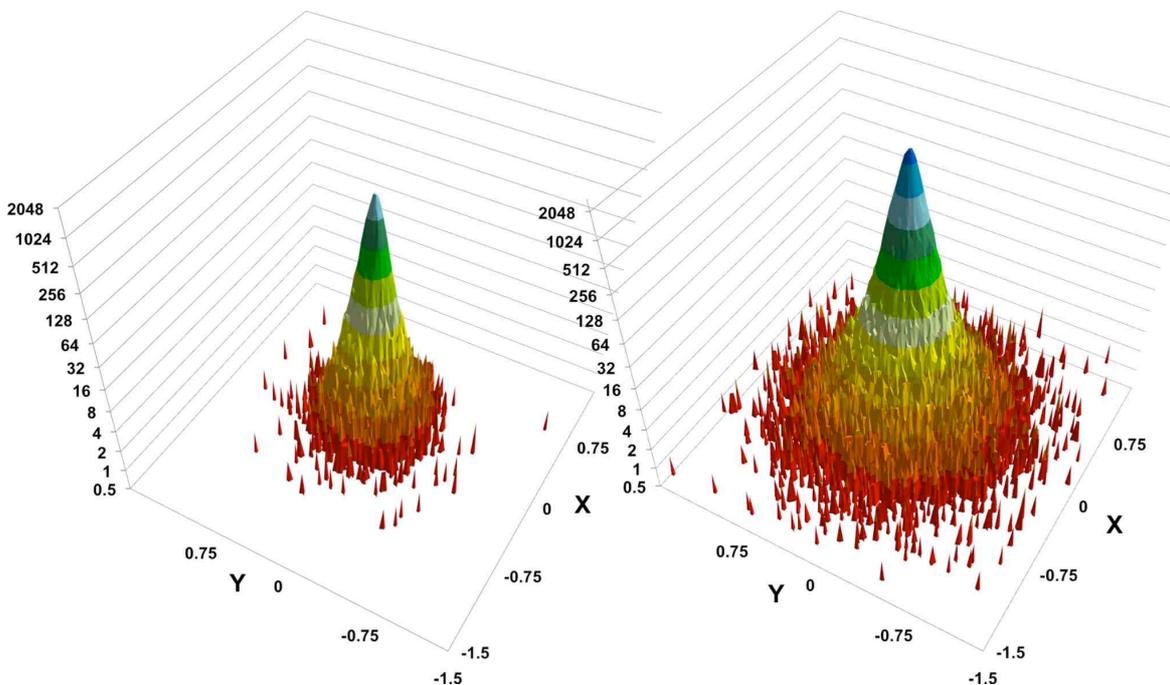

Figure 13. Simulation of the XY distribution of the impact points on the silicon detector, separately for alphas (left) and tritons (right), when the 16μm "thick" converter was irradiated in its center by a pencil neutron beam parallel to the z-axis.

In Figure 14 we show the simulated energy spectrum on the silicon detector for the flood irradiation, showing the separate contributions of alphas and tritons. The contribution of alphas is rather small, and it becomes smaller and smaller as the energy threshold increases: with a threshold at 1.5 MeV the alpha/triton yield ratio is ≈ 4.5%. Figure 15 shows again the deposited energy in the flood simulation, as compared with the isotropic simulation and with experimental data collected with the moderated AmBe neutron source. The energy calibration of the experimental spectrum was done by means of the known alpha and triton upper endpoint energy values, guided by the plots of Figure 14. The experimental spectrum had to be rescaled in order to compare it to the simulation results, and the scaling factor was chosen in such a way to have the same value at the alpha endpoint. The statistical error bars in the flood spectrum were tiny and were omitted for clarity, and the corresponding spectrum was reported as a continuous line. Like in the "thin" converter case the isotropic spectrum has a lower statistics and therefore larger error bars, due to the smaller number of neutrons hitting the converter. Again, it was normalized to the number of impinging neutrons without further rescaling, and as expected it is in good agreement with the flood spectrum. The choice of the horizontal error bar width was done like in the "thin" converter case.

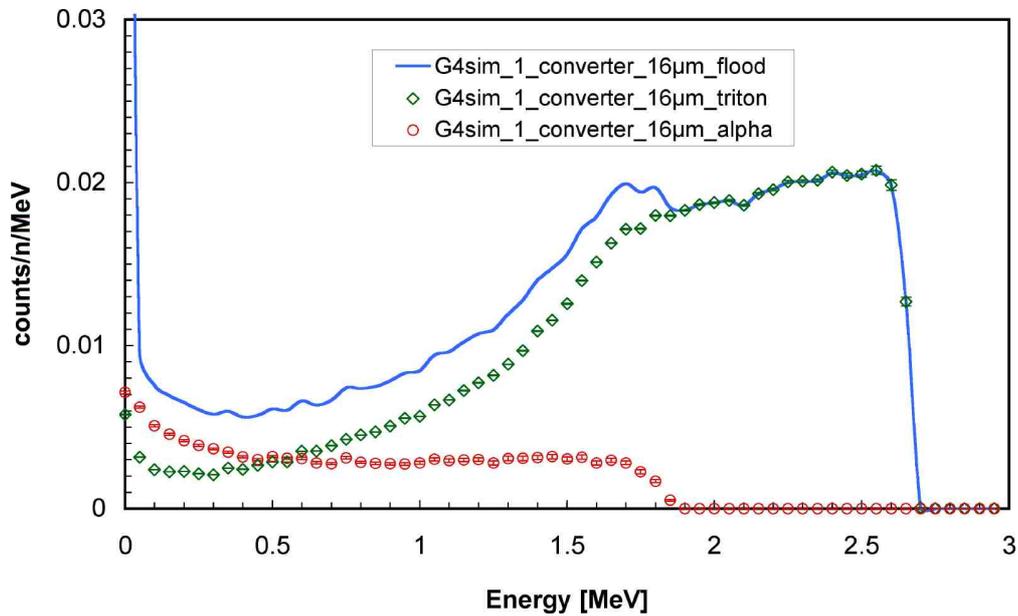

Figure 14. The simulated energy spectrum on the silicon detector for the flood irradiation in the "thick" 16 μm $^6$LiF converter configuration (Figure 1b), showing the separate contributions due to alphas and tritons.

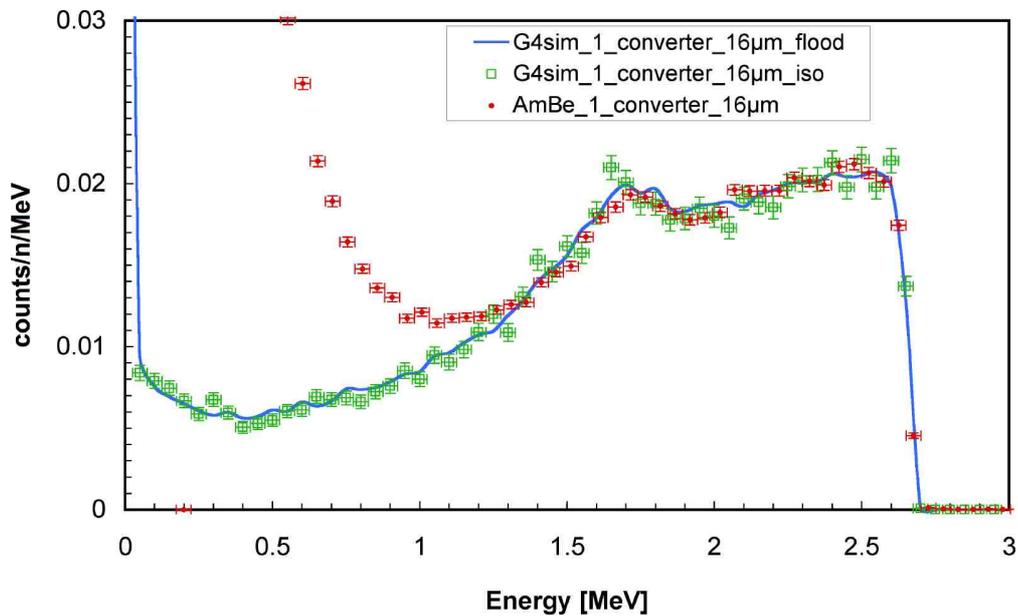

Figure 15. The energy spectrum on the silicon detector for the flood and isotropic simulations in the "thick" 16 μm $^6$LiF converter configuration (Figure 1b), compared with an experimental result obtained with a thermalized AmBe neutron source.

## 5   Thick neutron converter: sandwich configuration

The sandwich configuration examined in this case has a 16 μm thick converter on each face of the silicon detector (Figure 1c). The sandwich was simulated in the flood and isotropic irradiation schemes, and the distribution of the z-direction cosines for alphas and tritons hitting the silicon is shown in Figure 16, where the isotropic data were rescaled to the same number of incoming neutrons. In the plot a forward-to-backward asimmetry can be immediately appreciated in the flood irradiation scheme for tritons, due to the ≈ 8% neutron beam attenuation while crossing the first converter, that reduces the number of neutrons available for interaction in the second converter (the attenuation in the silicon detector is much lower, as the reaction cross section for thermal neutrons in silicon is several orders of magnitude smaller than in $^6$Li). The forward-to-backward ratio calculated over the full angular range indicates that the asymmetry is present also for alphas, as listed in Table 3, even though only ≈ 3% due to the useful alpha emission coming from a region closer to the converter surface facing the silicon detector.

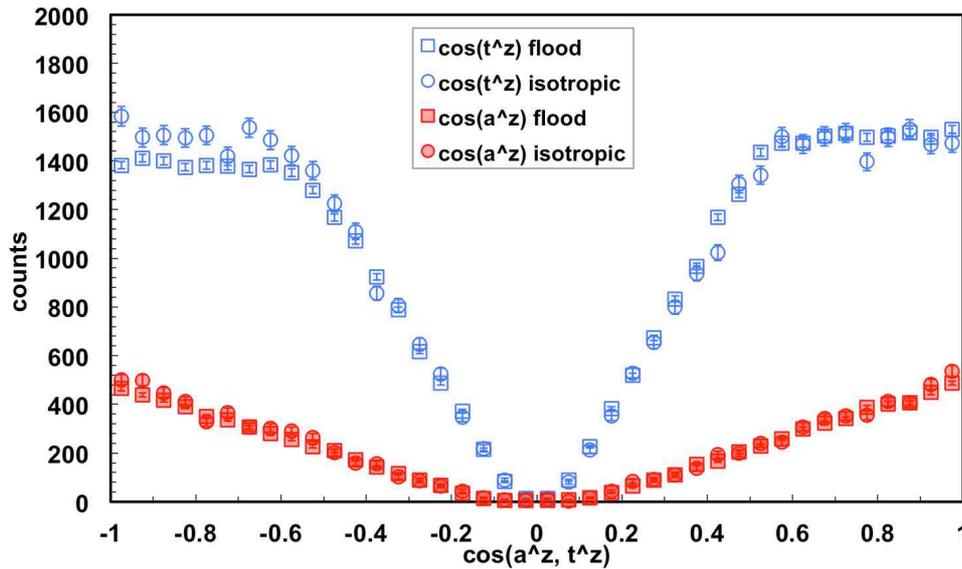

Figure 16. The z-direction cosines for alphas and tritons hitting the silicon detector in the sandwich configuration (flood and isotropic irradiation). A forward-to-backward asimmetry for tritons can easily be spotted, due to the neutron beam attenuation while crossing the first converter and the silicon detector.

|  | forward/backward ratio from 2x16µm $^6$LiF into silicon | | | |
|---|---|---|---|---|
|  | flood | deviation from 1 | isotropic | deviation from 1 |
| alpha | 1.027 ± 0.009 | 3σ | 1.00 ± 0.02 | < 1σ |
| triton | 1.083 ± 0.004 | 20.8σ | 1.00 ± 0.01 | < 1σ |

Table 3. The forward-to-backward ratio (counts in [0,1] divided by counts in [-1,0] from data of Figure 16) for alphas and tritons detected in silicon, in the cases of flood and isotropic irradiation of the sandwich configuration. The flood data show an asimmetry due to neutron absorpion in the first converter.

The energy spectra for the flood and isotropic simulations, as compared with the experimental data taken with the thermalized AmBe neutron source and with a (mostly) thermal neutron beam at the INES facility, are shown in Figure 17. The meaning of the error bars and the normalization are the same as for the previous cases of Figure 8 and Figure 15. The neutron source data should mimic the isotropic irradiation, whereas the beam data should mimic the flood irradiation. What we observe, indeed, is that there is no statistically significant difference between the energy spectra in the flood and isotropic irradiation schemes, and that the agreement between the simulations and the experimental data is remarkable (apart, of course, from the gamma ray background in the lower energy region).

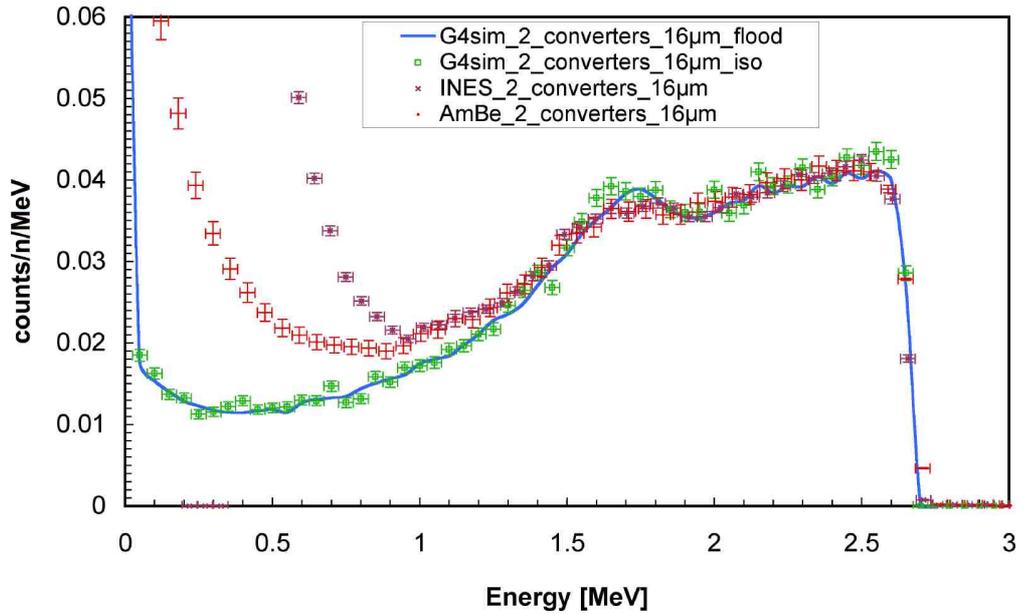

Figure 17. The energy spectrum on the silicon detector for the flood and isotropic simulations in the double side 16 μm $^6$LiF converter configuration (Figure 1c), compared with two experimental results obtained respectively with a thermalized AmBe neutron source and with a thermal neutron beam at the INES facility.

## 6   Thick neutron converter: double sandwich configuration

The configuration examined in this case was a double sandwich made of two units of the previously described sandwich stacked together (Figure 1d) and connected in parallel, thus producing a single OR-ed signal. The sandwich was simulated in the flood and isotropic irradiation schemes. In the flood irradiation each converter layer sees a reduced number of neutrons, due to capture and scattering interactions in the previous layer. As a consequence the second silicon detector was expected to show a smaller number of counts. This is exactly what was observed, as can be seen in Figure 18, where we show the deposited energy spectrum along with the individual contributions from the two silicon detectors. The contribution of the second silicon is 80% of the first one. In the isotropic irradiation scheme we did not expect differences between the two detectors in the stack, due to the complete simmetry of the system. Indeed this is exactly what happens, as shown in Figure 19.

The energy spectra for the flood and isotropic simulations, as compared with the experimental data taken with the thermalized AmBe neutron source and with a (mostly) thermal neutron beam at the INES facility, are shown in Figure 20. Also in this case the neutron source data should mimick the isotropic irradiation, whereas the beam data should mimick the flood irradiation. We observe again that there is no statistically significant difference between the energy spectra in the flood and isotropic irradiation schemes, and that the agreement between the simulations and the experimental data is remarkable (apart from the gamma ray background in the lower energy region).

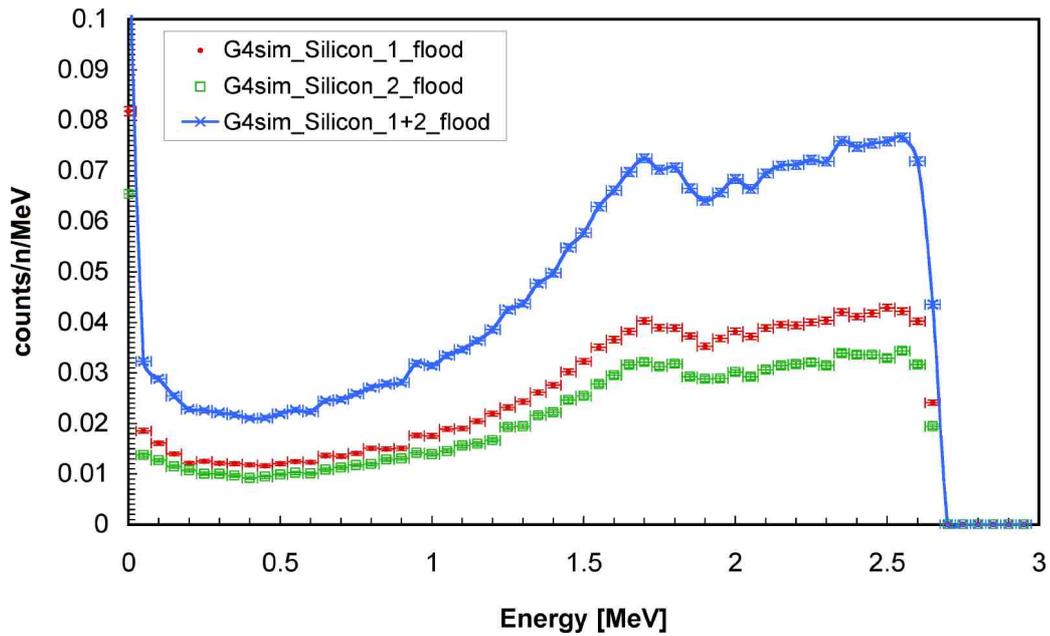
Figure 18. The energy spectrum on the two silicon detectors for the flood irradiation simulations in the double sandwich configuration (Figure 1d). Shown is also the sum spectrum, which is to be compared to the experimental data.

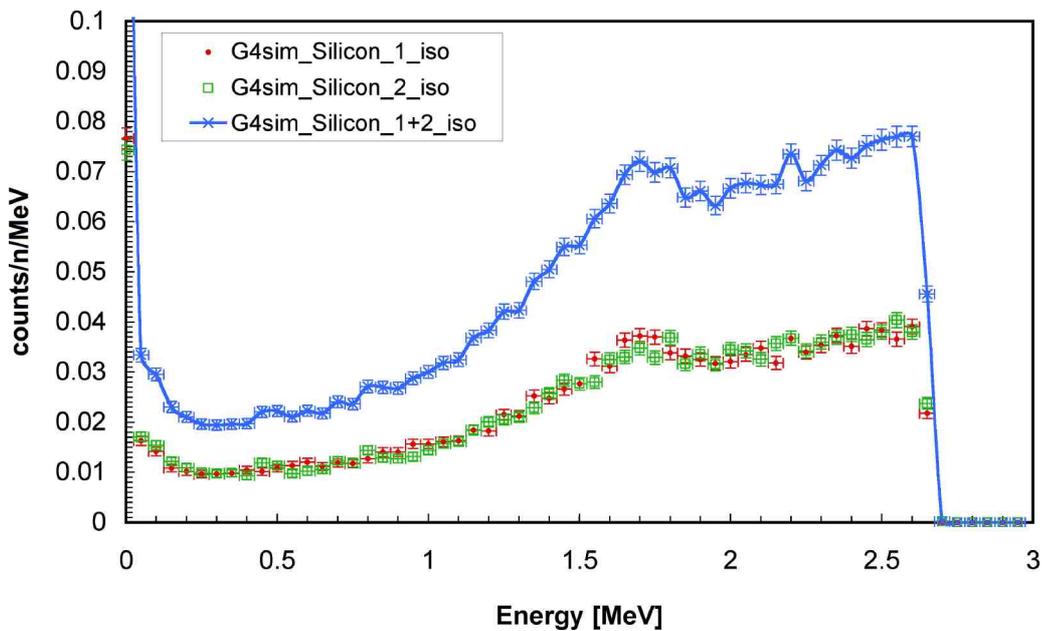
Figure 19. The energy spectrum on the two silicon detectors for the isotropic irradiation simulations in the double sandwich configuration (Figure 1d). Shown is also the sum spectrum, which is to be compared to the experimental data.

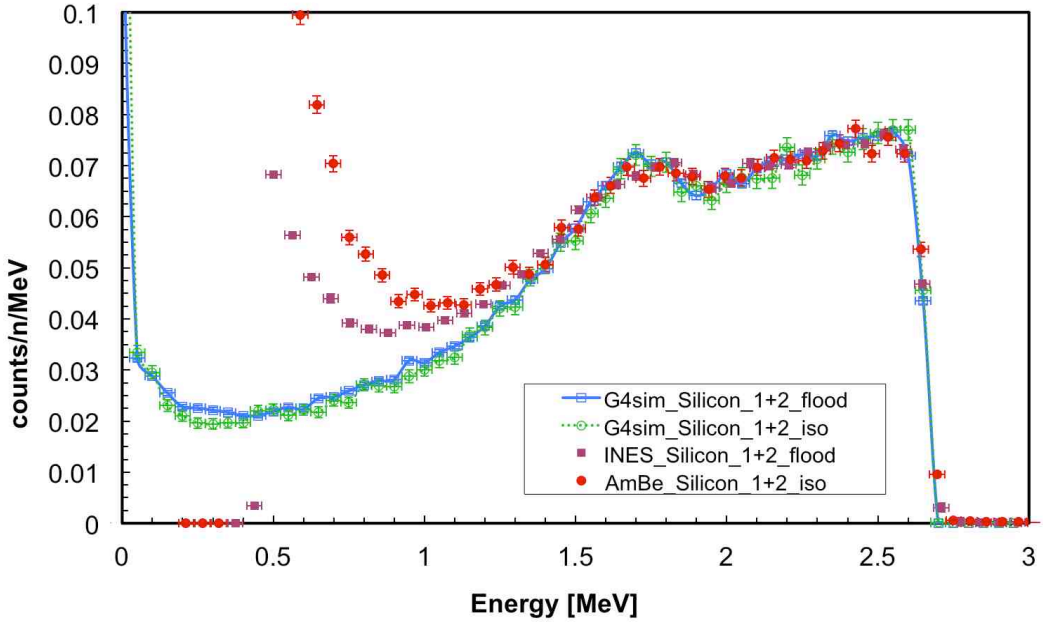

Figure 20. The energy spectrum on the silicon detector for the flood and isotropic simulations in the double sandwich configuration (Figure 1d), compared with two experimental results obtained respectively with a thermalized AmBe neutron source and with a thermal neutron beam at the INES facility.

## 7 Response to gamma rays

An important contribution to the energy spectrum of the detector comes from gamma rays, as clearly visible in the experimental data of Figure 8, Figure 15 and Figure 17. Disentangling and simulating the exact spectral shape of the gamma ray contribution is not realistic, as it strongly depends on the experimental conditions including the surrounding materials. Therefore we investigated the response to monoenergetic gamma rays, in the detector configuration with two 16μm converters on carbon fiber substrates. We performed nine simulations of $10^6$ monoenergetic gamma rays, with energies ranging from 0.5 to 4.5 MeV and a 0.5 MeV step. The corresponding deposited energy spectra are reported in Figure 21.

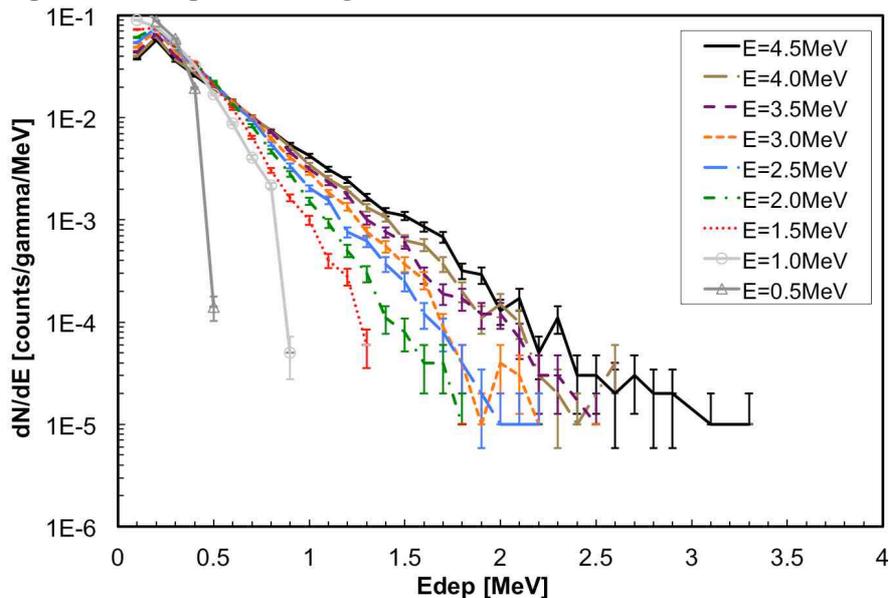

Figure 21. The simulated deposited energy spectrum on the silicon detector for irradiation with monoenergetic gamma rays ranging from 0.5 to 4.5 MeV.

By integrating the energy spectra above predefined threshold values, we computed the gamma ray sensitivity of the detector (i.e. probability that a gamma is detected as a neutron) for irradiation with monoenergetic gamma rays as a function of the detection threshold (Figure 22). As the minimum threshold investigated was 1 MeV, the sensitivity for gamma rays of 0.5 and 1.0 MeV resulted < $10^{-6}$ (no counts were detected, as the maximum possible energy release in the silicon is below threshold anyhow). As reference examples, from Figure 22 one can see that with a threshold set at 1.5 MeV the probability to detect a gamma ray of 2 MeV as a neutron is < $2 \cdot 10^{-5}$, whereas for a 4.5 MeV gamma ray the probability is ≈ $4 \cdot 10^{-4}$. Usually this figure of merit is evaluated with gamma rays from $^{60}$Co [1], whose maximum possible energy release is 1.3 MeV. Therefore in this case the only possibility to produce a signal above 1.5 MeV is from gamma pile-up, which thus implies a probability of the order of ≤$10^{-12}$.

From the plots in Figure 21 one can assume that the response of the detector to monoenergetic gamma rays is roughly exponential, therefore we played the following simple but useful consistency check. A trendline of the form $n(E) = a \cdot e^{-kE}$ was drawn through the data points, and an indicative value of $k$ was extracted for each spectrum (excluding 0.5 and 1.0 MeV). Then the same procedure was performed on the experimental spectra related to the neutron source (AmBe) and to the neutron beam (INES), only considering the low energy data points between 0.1 and 0.7 MeV, to exclude the strong 59 keV gamma ray contribution from the $^{241}$Am in the source and the onset of the dominant triton contribution. The results are listed in Table 4 and plotted in Figure 23, where we extrapolated the hypothetical corresponding gamma ray energy range for the experimental data. The 4.2 MeV value obtained for the AmBe source sounds quite reasonable in light of the 4.4 MeV gamma rays it emits, and the 2.15 MeV obtained for the thermal neutron beam data sounds reasonable as well, in light of the gamma ray background basically due to neutron capture in the moderator [basically 2.2 MeV from H(n,γ)D] and activation of materials around the detector in the experimental hall.

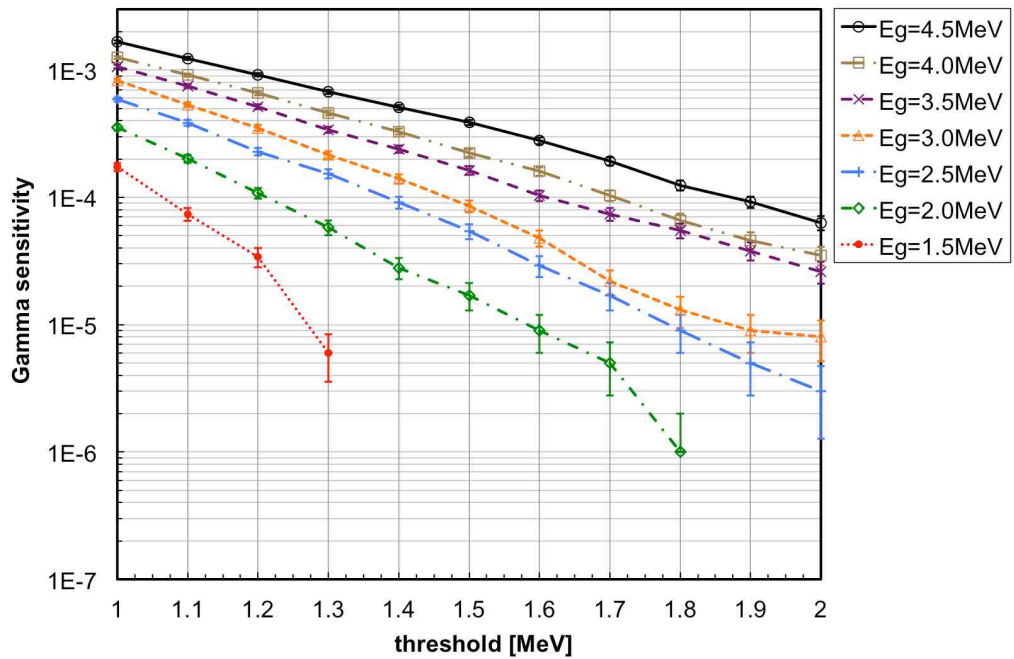

Figure 22. The gamma sensitivity (i.e. detection probability per incident gamma) of the detector for irradiation with monoenergetic gamma rays as a function of the detection threshold. As the minimum threshold investigated was 1 MeV, the sensitivity for gamma rays of 0.5 and 1.0 MeV resulted < $10^{-6}$.

| Eg [MeV] | exponential slope $k$ [MeV$^{-1}$] |
|---|---|
| 4.5 | 3 |
| 4 | 3.5 |
| 3.5 | 3.8 |
| 3 | 4.3 |
| 2.5 | 4.6 |
| 2 | 5.2 |
| 1.5 | 6.4 |
| **INES 2.15** | 5 |
| **AmBe 4.2** | 3.3 |

Table 4. The exponential slope parameter of the deposited energy spectrum for the simulated monoenergetic gamma rays, as estimated from trendlines of the form $n(E) = a \cdot e^{-kE}$ to the data of Figure 21. A similar estimate from the low energy part of the experimental spectra in the INES and AmBe cases gives *k ≈ 5* and *k ≈ 3.3*, roughly corresponding to monoenergetic gamma rays of ≈ 2.15 and ≈ 4.2 MeV.

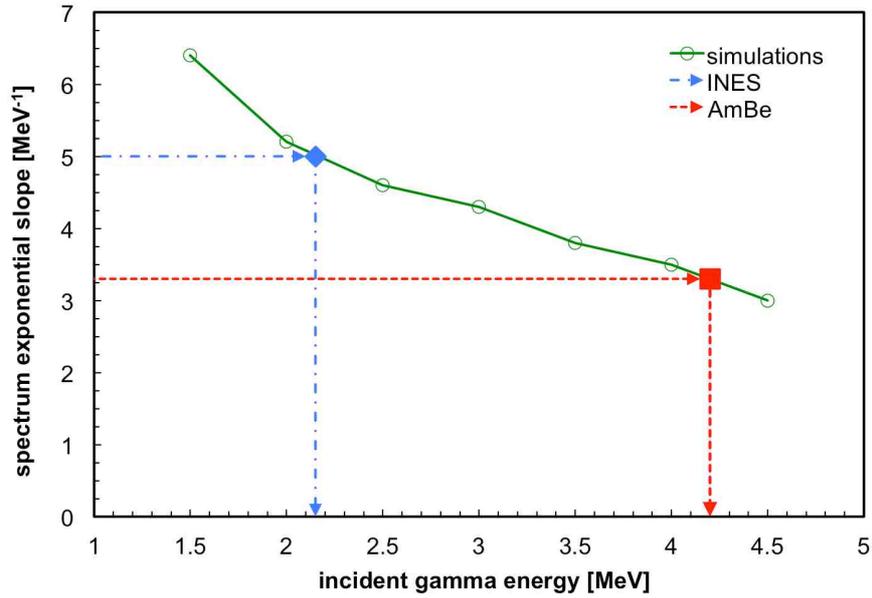

Figure 23. The exponential slope parameters of Table 4 as a function of the corresponding monoenergetic gamma ray energy. Shown are the two extrapolated points for the neutron source (AmBe) and neutron beam (INES) gamma ray contributions.

We also evaluated the background contribution due to secondary particles, mainly produced in the converter, in the carbon fiber substrate and in the detector itself. Table 5 lists the number of electrons, gamma rays and protons+ions, excluding alphas and tritons, hitting the silicon detector per incident neutron, which in light of the already tiny efficiency can be neglected.

| e- | | gamma | | protons+ions | |
|---|---|---|---|---|---|
| all | E > 1.5MeV | all | E > 1.5MeV | all | E > 1.5MeV |
| 8.4·10$^{-4}$ | 1.1·10$^{-5}$ | 1.2·10$^{-3}$ | 1.0·10$^{-3}$ | 7·10$^{-6}$ | 0 |

Table 5 – Number of secondary particles hitting the silicon per incident neutron (excluding alphas and tritons)

## 8 Discussion

The data plotted in Figure 8, Figure 15, Figure 17 and Figure 20 allow to evaluate the neutron detection efficiency respectively for the four detector configurations shown in Figure 1a-d. Indeed, by integrating the simulated spectra from several threshold values upward we obtained the expected detection efficiency as a function of the threshold itself for each configuration. The results are shown in Figure 24, Figure 25, Figure 26 and Figure 27. The same procedure, applied to the experimental data, provides the measured detection efficiency. We remark that the error bars (statistical) are within the symbols, and that the agreement between the simulations and the real data is very good apart from the points at lower threshold where the contribution from gamma rays was still relevant. We remark, indeed, that the experimental data can be affected by a systematical error due to the normalization, as the real neutron flux in the experimental conditions was not known. A good tradeoff between efficiency and purity can be assumed with 1.5 MeV threshold value, where the sensitivity to the typical activation gamma rays is about $10^{-5}$ whereas the sensitivity to the reference $^{60}$Co is of the order of $\leq 10^{-12}$ (Figure 22). Above that threshold there is no difference between the simulated and experimental efficiency, and this holds for both the flood and isotropic simulated irradiation schemes and for both the source and beam experimental conditions. This is a remarkable result, as it testifies the uniform behavior of the detector under different irradiation conditions.

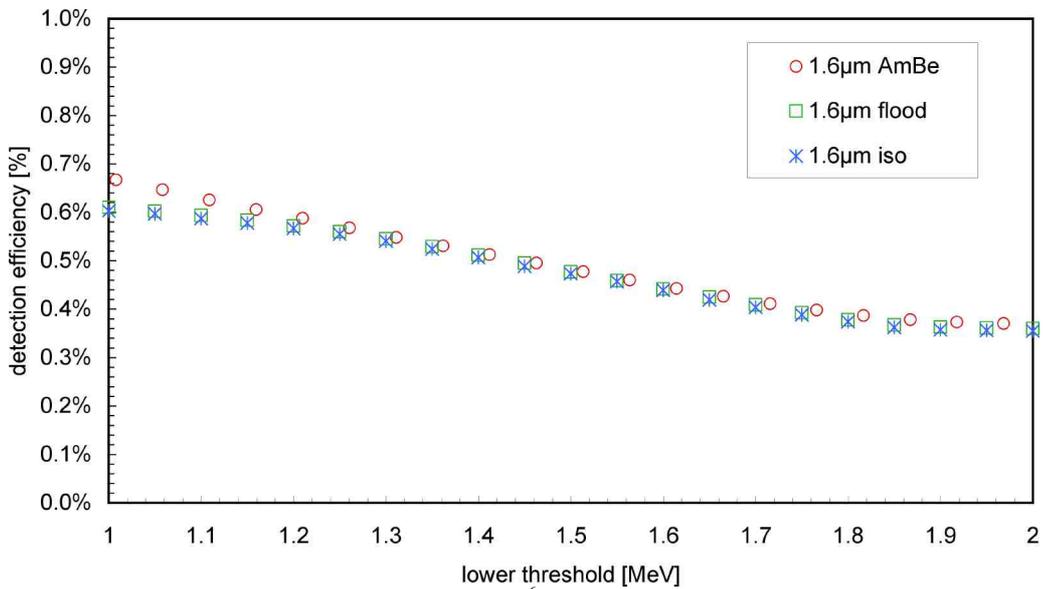

Figure 24. Neutron detection efficiency of the "thin" 1.6 μm $^6$LiF detector configuration as a function of the detection threshold, for the AmBe experimental data and the simulated flood and isotropic iradiations.

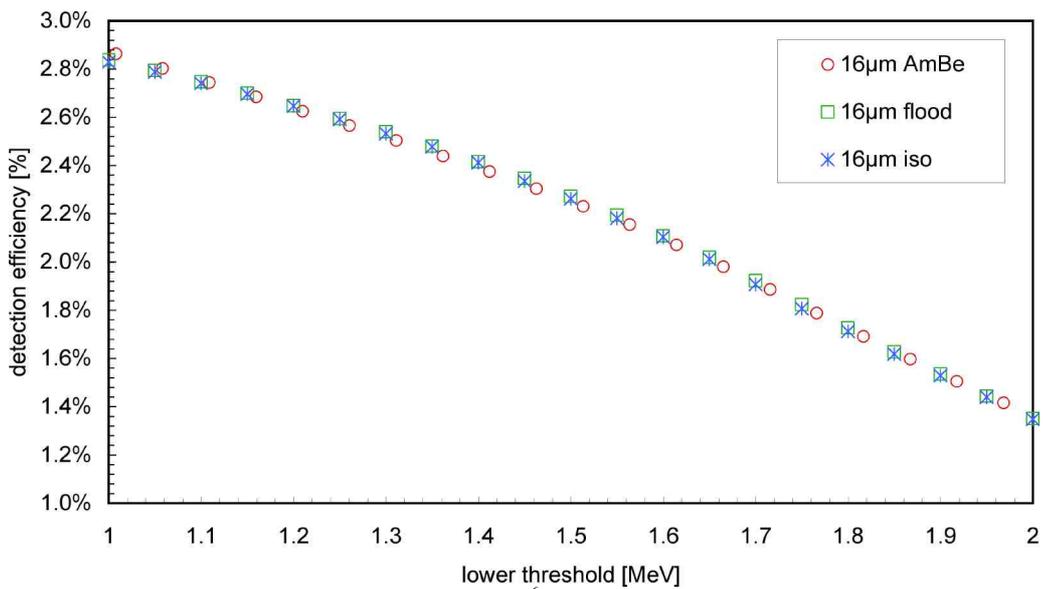

Figure 25. Neutron detection efficiency of the "thick" 16 μm $^6$LiF detector configuration as a function of the detection threshold, for the AmBe experimental data and the simulated flood and isotropic iradiations.

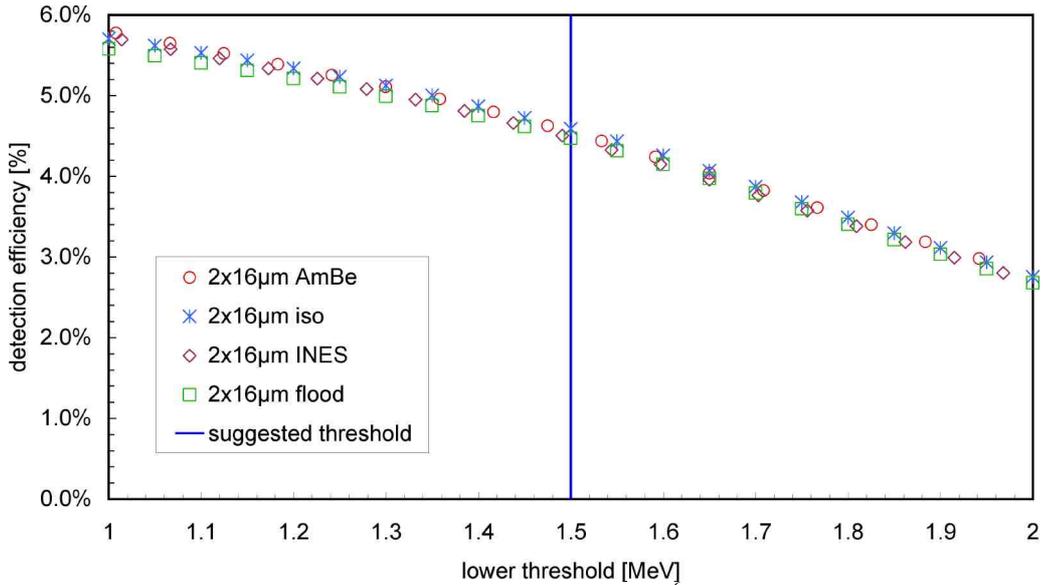

Figure 26. Neutron detection efficiency of the double side "thick" 16 μm $^6$LiF detector configuration as a function of the detection threshold, for the AmBe and INES experimental data, and the simulated flood and isotropic iradiations.

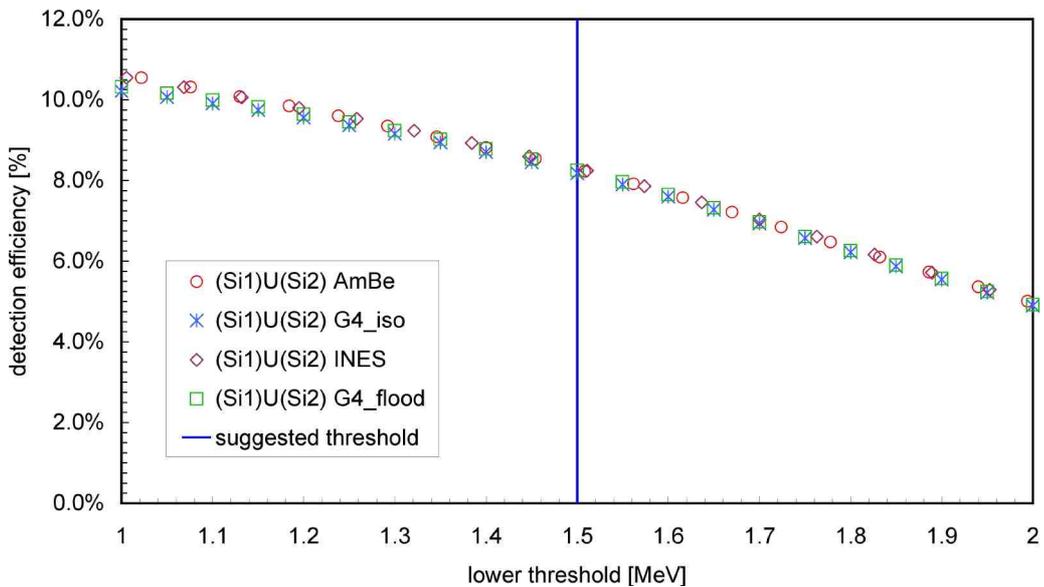

Figure 27. Neutron detection efficiency of the double sandwich detector configuration as a function of the detection threshold, for the AmBe and INES experimental data, and the simulated flood and isotropic iradiations.

As for the gamma/neutron discrimination performance, we calculated the ratio between the simulated gamma sensitivity (Figure 22) and the neutron efficiency of the double side "thick" 16 μm $^6$LiF detector (Figure 26), as a function of the threshold level for several monoenergetic gamma rays. This essentially provides the fraction of counts in the spectrum due to gamma rays but interpreted as neutrons when the detector is subject to an equal flux of thermal neutrons and (monoenergetic) gammas, i.e. the gamma-to-neutron contamination probability. The resulting plot, shown in Figure 28, indicates that with the threshold at the suggested value of 1.5 MeV the contamination probability from typical activation gamma rays ($\approx \leq 2$ MeV) is of the order of $10^{-4}$. The gamma-to-neutron contamination from $^{60}$Co is $\approx \leq 10^{-11}$. The gamma sensitivity features of the double sandwich configuration are basically the same, because by doubling the detectors we double both the neutron and the gamma sensitivity.

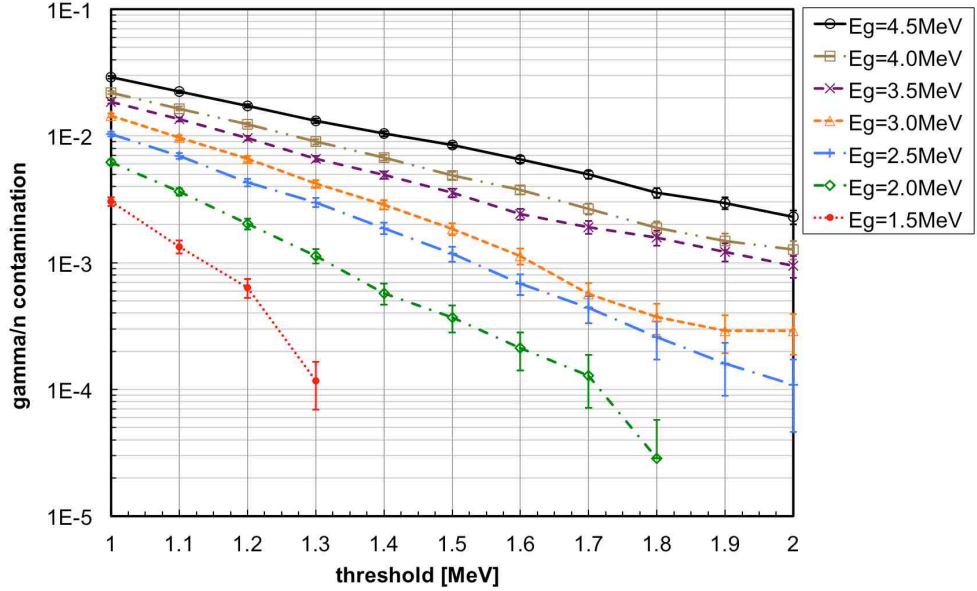

Figure 28. The gamma/neutron contamination probability for equal fluxes (i.e. the ratio of the gamma sensitivity to the neutron efficiency) as a function of the detection threshold for the double side "thick" 16 μm $^6$LiF detector.

## 9 Conclusion

In the framework of the worldwide research and development programs seeking new technologies and methods for the neutron detection, we have investigated a possible alternative featuring a reasonable efficiency and an excellent gamma/n rejection. We are aware that such a solution is not suitable for every application, especially in cases where large area detectors are required, but nevertheless it could pave the way to the implementation of small to medium size systems useful as portable and/or environmental monitors. Its main disadvantage being the limited size of the typical semiconductor detectors, this solution provides however several advantages: simple availability of the components, along with the simple mechanical/electronic setup, low voltage operation in air with no need of special gases, uniform efficiency and response across its sensitive area due to the planar geometry, non-permanent assembly as the detector and the neutron converters can be individually replaced. The comparison of the experimental data with the simulation results looks very encouraging, and the detection efficiency is of the order of 5% for the double converter configuration and of 10% for the double sandwich stacked configuration. The secondary reactions occurring in the detector itself can be neglected, whereas the gamma/neutron discrimination has been studied and quantified resulting $\leq 10^{-12}$ around the $^{60}$Co energy.

Several applications of the proposed detection technique are already in use, like for instance at the n-TOF spallation neutron beam facility, and others are under development for homeland security and nuclear safeguards. Our future plans concern the possible application in a distributed monitoring system for spent fuel interim storage sites, where many thermal neutron detectors would be installed around the casks to keep the outcoming neutron flux under control, in order to detect any sign of abnormal behavior due to safety and/or security issues. To this purpose our next step will be to validate the simulations with an absolute efficiency measurement to be performed by means of a calibrated thermal neutron field, possibly in a metrology facility.

## 10 Acknowledgments

We are indebted to Carmelo Marchetta and Salvatore Russo at INFN-LNS Catania for the preparation of the $^6$LiF converters and for the help with the AmBe source issues, Alfio Pappalardo now at ELI-NP Bucharest for his invaluable contribution with the detector assemblies, Cirino Vasi at CNR-IPCF Messina for providing us access to the INES facility at Rutherford Appleton Laboratory.